\documentclass[11pt,a4paper]{article}
\usepackage{jheppub,subfig,graphicx,amsmath}
\usepackage{bbold}
\usepackage{multirow}
\usepackage[bottom]{footmisc}
\usepackage{graphicx,booktabs,array,float}
\newcommand{\be}{\begin{equation}}
\newcommand{\ee}{\end{equation}} 
\newcommand{\ba}{\begin{array}}
\newcommand{\ea}{\end{array}}
\newcommand{\bea}{\begin{eqnarray}}
\newcommand{\eea}{\end{eqnarray}}
\newcommand{\tD}{{\tilde D}}
\newcommand{\tA}{{\tilde A}}
\newcommand{\tz}{{\tilde \zeta}}
\newcommand{\cF}{\mathcal{F}}
\newcommand{\cG}{\mathcal{G}}
\newcommand{\bG}{\mathbb{G}}

\title{Flow Equation and Fermion Gap in the Holographic Superconductors}
\author{Taewon Yuk and Sang-Jin Sin}
\affiliation{Department of Physics, Hanyang University, Seoul 04763, South Korea}
\emailAdd{tae1yuk@gmail.com, sangjin.sin@gmail.com}
\abstract
{We reconsider the fermion spectral function in the presence of the Cooper pair condensation and identified the interaction type of complex scalar and fermion, which gives consistent results with the expected s-wave superconductor for the first time. We derive the matrix Riccati equation, which allows the precise calculation of the fermion spectral function. Apart from the gap structure, we studied the effect of the chemical potential and the density and compared it with the BCS theory. We found that two theories give similar results in small chemical potential but very different ones in the high-density case, which we attribute to the correlation effect. 
}
\arxivnumber{arxiv number}

\begin{document}
\maketitle
	
	\section{Introduction}
	Understanding the high Tc superconductivity has been one of the major motivations to study strongly correlated systems. Therefore there have been huge activities for holographic superconductivity \cite{Gubser:2008px, Hart2008, Horo2009, Siop2010, siopsis2011holographic, Horowitz:2009ij, Sin:2009wi, Pan:2012jf, Brihaye:2010mr, banerjee2013holographic, Ishii:2012hw, Choun2021a} based on the gauge gravity duality \cite{Maldacena:1997re, Witten:1998qj, Gubser:1998bc, Hartnoll:2016apf}. 
	The novel achievement was to construct a new dynamical mechanism of the instability to give the U(1) symmetry breaking from the dynamics of the scalar-vector-gravity system\cite{Gubser:2008px, Hart2008}. Despite the similarity of the abelian Higgs Model in flat space, holographic models need not use the Higgs potential. Instead, it can rely on the horizon instability under cooling the system, leading to the transition from a hairless to a hairy black hole where some of the charges are sent out to the black hole exterior as a scalar field condensation.

	After the original works\cite{Gubser:2008px, Hart2008}, which were for s-wave superconductivity, p-wave and d-wav superconductivity were also studied using the vector and spin-2 tensor fields \cite{Gubser:2008wv,Ammon:2010pg,Chen:2010mk,benini2011holographic,Kim:2013oba} and the spectrum of the fermions under the presence of the hair was studied in \cite{Faulkner:2009am} for the s-wave and also in \cite{Gubser:2010dm}, \cite{Chen:2011ny,benini2011holographic} for p- and d-waves. 
	
	However, to our surprise, 
the s-wave gap structure in the fermion spectrum under the presence of the complex scalar condensation has not been shown clearly until today. Authors of \cite{Faulkner:2009am} constructed some of the most natural-looking interaction terms between the fermion and complex scalar, 
 \begin{align}
 	{\cal L} =\Phi {\bar\psi}_{c}\Gamma\psi, \quad \label{Lint} \hbox{ with} 
 	\Gamma=1, \; \Gamma^{5}.
 \end{align}
 Such scalar-Majorana-mass interaction term is the most anticipated analog of the flat space BCS theory. However, the resulting spectrum is very different: spectral function does not have a clear gap and has high weights in the regime where one does not expect any. 

In this paper, we reconsider the problem of constructing the fermion spectral function with the s-wave gap in the presence of the complex scalar. We find that apart from the interaction type mentioned above, two other terms can be considered as scalars, and one of them generates the desired superconducting gap with expected features. These are interactions of the type \eqref{Lint} but with
\begin{align}
	\Gamma=\Gamma^{ {z}}, \quad \Gamma^{z}\Gamma^{5},
\end{align}
which are vectors from the bulk point of view but classified as scalars from the boundary view. {\it It turns out that $\Gamma^{z}$ gives the gap with the features expected in the superconductivity.}

 Another issue is to find a flow equation systematically. In the numerical computation of the fermion spectrum, the most crucial source of the error comes from the boundary condition at the black horizon, where the fluctuating spinor solutions are singular, forcing us to impose the boundary condition of the horizon. To avoid the problem, one need to extend the flow equation method \cite{Liu:2009dm}, where the Green function itself is the variable and regular at the horizon. Such regularity enables us to calculate the precise solution much shorter time. However, incorrect flow equation seems to be an origin of the incorrect spectral shape. 

We found that our spectral function gives the gap structure expected from the superconductivity. In detail, however, the correlation effect introduces a few differences.
We also studied the effect of the chemical potential and density and compared it with the BCS theory. We found that the two theories are similar in the small chemical potential regime but are very different in the high-density case. We suggest that this is due to the strong correlation between the electrons.
	 
	\section{Holographic Superconductivity vs. BCS Theory}
	 The action of our model is given by
	\begin{align}
		S_{tot}=&S_{\psi}+S_{int}+S_{bdy}+S_{g,A,\Phi }, \label{eq:actioni} \\
		S_{\psi}=&\int d^4x\sqrt{-g}\left[i\bar{\psi}(\Gamma^{\mu}D_{\mu}-m)\psi-i\bar{\psi_c}(\Gamma^{\mu}D_{\mu}^*-m)\psi_c\right], \label{eq:bulkaction} \\
		S_{int}=& \int d^4x \sqrt{-g} \left(\Phi \bar{\psi}\Gamma^{z}\psi_c+h.c. \right), \label{interaction}\\
		S_{bdy}=& i\int d^3x\sqrt{-h}(\bar{\psi}\psi+\bar{\psi_c}\psi_c) , \label{eq:actionf} \\
		S_{g,A,\Phi }=& \int d^4x\sqrt{-g} \left(R+\frac{6}{L^2}-\frac{1}{4}F^2_{\mu\nu} -|\mathcal{D}_{\mu}\Phi |^2 -m^{2}_{\Phi}|\Phi |^{2}\right),
	\end{align}
	where $g_{\mu\nu}$, $A_\mu$ and $\Phi $ are considered as the background fields. 
	In this paper, we do not consider the back reaction effects. 
	Our main goal is to show that our model reproduces the qualitative features of the fermion spectrum of BCS theory in the s-wave case. Therefore we first want to ensure that our model is comparable to the BCS theory at the level of the degrees of freedom. The point in comparison is that half of the bulk degrees of freedom in the holographic theory are projected out, and the residual physical ones are the boundary behavior of the surviving ones. In terms of the two-component spinors $\psi_{\pm}$ with $\psi^{T}=(\psi_+, \psi_- )$, the interaction term can be written as 
	\begin{align}
		S_{int}=\int d^4x \sqrt{-\tilde{g}}[\Phi^* \psi_-^T(i\sigma_2)\psi_+ +\Phi \psi_+^{\dagger}(-i\sigma_2)\psi_-^*+h.c.],
	\end{align}	
	 where $\psi_c=\psi^*$ and $\sqrt{-\tilde{g}}=\sqrt{g_{xx}g_{yy}}$. The first two terms of $S_{int}$ can be interpreted as the source term and its hermitian conjugate as the response term; it will be seen later.
	 Since the boundary degrees of freedom correspond to source parts, we request $\psi_+, \psi_-^*$ to match the fermion action of BCS theory. We introduce the analog of the Nambu-Gork'ov spinor $\Psi=(\psi_+, \psi_-^*)$ to rewrite the action as 
	\begin{align}
		S_{int}=\int d^4x \sqrt{-\tilde{g}}\left[\Psi^{\dagger} 
		\begin{pmatrix}0&-\Phi i\sigma_{2}\\
		\Phi^{*} i\sigma_{2}&0
		\end{pmatrix}
		 \Psi +h.c.\right].\label{Nambu}
	\end{align} 
	Furthermore, if we introduce the $c_{\uparrow\downarrow}$ as the components of $\psi_+, \psi_-^*$, i.e, 
	 	\begin{align}
		\psi_+=&(c_\uparrow(z,\omega, k),c_\downarrow(z,\omega, k))^T, \nonumber \\
		\psi_-^*=&(c_\uparrow^*(z,-\omega, -k),c_\downarrow^*(z,-\omega, -k))^T,
	\end{align}
	the interaction term can be written in terms of the components as follows.
	\begin{align}
		S_{int}=&\int dzd^3k \sqrt{-\tilde{g}}[\Phi (-c_\uparrow^*(z,\omega,k)c_\downarrow^*(z,-\omega,-k)+c_\downarrow^*(z,\omega,k)c_\uparrow^*(z,-\omega,-k)) \nonumber \\
		 +&\Phi^*_c(c_\uparrow(z,-\omega,-k)c_\downarrow(z,\omega,k)-c_\downarrow(z,-\omega,-k)c_\uparrow(z,\omega,k))] \label{eq:interaction}
	\end{align}
	If we compare this with the action of the BCS theory\cite{Faulkner:2009am} 
	\begin{align}
		&S[c]=\int d^{3}k [c_\alpha^{\dagger}(\omega,k)(i\omega-\epsilon_k)c_\alpha(\omega,k)-\Delta(k)c_{\uparrow}^{\dagger}(\omega,k)c_{\downarrow}^{\dagger}(-\omega,-k) \nonumber \\
		&\quad\quad\quad\quad -\Delta^*(k)c_{\uparrow}(\omega,k)c_{\downarrow}(-\omega,-k)] ,\label{eq:bcs}
	\end{align}
	we find that with the correspondence 
	\begin{align}
		c_{\alpha} (z,\omega,k) \leftrightarrow c_{\alpha} (\omega,k),\quad 2\Phi (z,k) \leftrightarrow \Delta(k), 
	\end{align}
	our system match with BCS theory at the level of degrees of freedom. Here $\alpha=\uparrow,\downarrow$ are spin indices and $\Delta$ is the superconducting order parameter and we used the anti-commute property of $c(\omega,k)$.

	\section{Flow Equation}
	In this section we develop the flow equation which is our main method to discuss the detailed gap structure. 
	\subsection{Dirac Equation}
	Gamma matrices and the bosonic fields are given by
	\begin{align}
		\Gamma^{\underline{t}}&=\sigma_1 \otimes i \sigma_2, \quad \Gamma^{\underline{x}}=\sigma_1 \otimes \sigma_1, \quad \Gamma^{\underline{y}}=\sigma_1 \otimes \sigma_3, \quad \Gamma^{\underline{z}}=\sigma_3 \otimes \sigma_0, \quad \Gamma^{\underline{5}}=\sigma_2 \otimes \sigma_0, \label{eq:rep} \\
		ds^2&=\frac{-f(z)\chi(z)}{z^2}dt^2+\frac{dx^2+dy^2}{z^2}+\frac{dz^2}{z^2f(z)}, \quad A_{\mu}=(A_t(z),0,0,0), \quad \Phi=\Phi(z),
		\label{eq:background}
	\end{align}
	where underlined indices represent tangent space indices. In our $\Gamma$ matrix representation, the charge conjugation operator is $C=\mathbb{1}K$, where $K$ is the complex conjugation. Now we can write the Dirac equation as follows.
	\begin{align}
		(\Gamma^{\mu} D_{\mu}-m)\psi-i\Phi \Gamma^{z}\psi_c=0,
	\end{align}
	where $D_\mu=\partial_\mu+\frac{1}{4}\omega_{\underline{\nu}\underline{\lambda},\mu}\Gamma^{\underline{\nu}\underline{\lambda}}-iq A_\mu$, $\omega_{\underline{\nu}\underline{\lambda},\mu}$ is the spin connection.
	Substituting 
	$$\psi=(-gg^{zz})^{-1/4} e^{-i\omega t+i k_xx+i k_yy}\zeta(z),$$ we can get simplified Dirac equation\cite{Liu:2009dm}:
	\begin{align}
		\left[\Gamma^{\underline{z}}\partial_z-{i}\left(\frac{(\omega+q A_t)}{f(z)\sqrt{\chi(z)}} \Gamma^{\underline{t}}-\frac{1}{\sqrt{f(z)}}(k_x \Gamma^{\underline{x}} +k_y \Gamma^{\underline{y}})\right)-\frac{m}{z\sqrt{f(z)}}\right] \zeta-i \Phi \Gamma^{\underline{z}} \zeta_{c}=0. \label{eq:diraceq}
	\end{align}
	The charge conjugation of this equation is given by
	\begin{align}
		\left[\Gamma^{\underline{z}}\partial_z+{i}\left(\frac{(\omega-q A_t)}{f(z)\sqrt{\chi(z)}} \Gamma^{\underline{t}}-\frac{1}{\sqrt{f(z)}}(k_x \Gamma^{\underline{x}} +k_y \Gamma^{\underline{y}})\right)-\frac{m}{z\sqrt{f(z)}}\right] \zeta_{c}+i \Phi ^* \Gamma^{\underline{z}} \zeta=0. \label{eq:cdiraceq}
	\end{align}
	
	\subsection{Determining Source and Condensation}
	For two-component spinor $A,B$, we introduce a notation $(A;B)=\begin{pmatrix}A\\ B\end{pmatrix}$, so that 4-component spinor $\psi=\begin{pmatrix}\psi_{+}\\ \psi_{-}\end{pmatrix}$ can be written as $\psi=(\psi_{+};\psi_{-})$. \\
	To see which components are the source, we need to investigate the variation of the total action. The variation of the bulk action is
	\begin{align}
		\delta S_{bulk}=EOM+ i\int d^3x (\bar{\zeta}\Gamma^{\underline{z}}\delta\zeta-\delta\bar{\zeta}\Gamma^{\underline{z}}\zeta-\bar{\zeta_c}\Gamma^{\underline{z}}\delta\zeta_c+\delta\bar{\zeta_c}\Gamma^{\underline{z}}\zeta_c), \label{eq:bulkv}
	\end{align}
	while the variation of boundary action is
	\begin{align}
		\delta S_{bdy} = i\int d^3x (\delta\bar{\zeta}\zeta+\bar{\zeta}\delta\zeta+\delta\bar{\zeta_c}\zeta_c+\bar{\zeta_c}\delta\zeta_c) .\label{eq:bdyv}
	\end{align}
	Adding \eqref{eq:bulkv} and \eqref{eq:bdyv} and expressing the result in terms of the two-component spinors,
	\begin{align}
		\delta S_{bulk}+\delta S_{bdy} =2i\int d^3x\Big(-\zeta_{-}^{\dagger}\sigma_1\delta\zeta_{+}+\zeta_{c+}^{\dagger}\sigma_1\delta\zeta_{c-}\Big)+h.c,
	\end{align}
	where $\zeta=(\zeta_+;\zeta_-)$ and $\zeta_c=(\zeta_{c+};\zeta_{c-})$. Now we see that we should choose $(\zeta_{+};\zeta_{c-})$ to be the source whose values are fixed on the AdS boundary to make the variation of the total action zero. This source definition is equivalent to that of Nambu-Gork'ov spinor in eq. \eqref{Nambu}. While to find the condensation, we only consider the boundary action. Because the contribution of the bulk action to the effective action is zero due to the equation of motion.
	\begin{align}
		S_{eff}&=S_{bdy}= i\int d^3x\sqrt{-h}(\bar{\psi}\psi+\bar{\psi_c}\psi_c) = i\int d^3x(\bar{\zeta}\zeta+\bar{\zeta_c}\zeta_c) \nonumber \\
		&=\int d^3x 
		\bar{\zeta}_-\zeta_+ +\bar{\zeta}_{c+}\zeta_{c-}+h.c,
	\end{align}
	where $\bar{\zeta}_-= {\zeta_-}^\dagger \gamma^t$ with $\gamma^t=-\sigma_2$. From this, we can choose $(\zeta_{-};\zeta_{c+})$ as the condensation which is the conjugate momentum of the source $(\zeta_{+};\zeta_{c-})$.
	
	\subsection{Derivation of the Flow Equation} 
	We define 4-components spinor $\xi^{(S)}$ and $\xi^{(C)}$ by the ordering of the source and condensation: 
	\begin{align}
		\xi^{(S)}=(\zeta_+ ; {\tilde\zeta}^*_-) , \quad \xi^{(C)}=(\zeta_-,{\tilde\zeta}_+^*). \label{eq:xidef}
	\end{align}
	Since we introduced $\zeta_{c}$ as the complex conjugation of $\tilde\zeta$, which is a copy of $\zeta$, $\tz$ should be treated as an independent field. If we rearrange (\ref{eq:diraceq}-\ref{eq:cdiraceq}), the two Dirac equations can be written as
	\begin{align}
		\partial_z\xi^{(S)}+\mathbb{M}_1\xi^{(S)}+\mathbb{M}_2\xi^{(C)}=0,\label{eq:xseq} \\
		\partial_z\xi^{(C)}+\mathbb{M}_3\xi^{(C)}+\mathbb{M}_4\xi^{(S)}=0, \label{eq:xceq}
	\end{align}
	where
	\begin{gather}
		\mathbb{M}_1=-\frac{m}{z\sqrt{f(z)}}
		\begin{pmatrix}
			\sigma_0 & 0 \\
			0 & -\sigma_0 
		\end{pmatrix}, \quad
		\mathbb{M}_3=-\mathbb{M}_1, \\
		\mathbb{M}_2=
		\begin{pmatrix}
			\mathbb{N}(q) & \mathbb{P} \\
			\mathbb{P}^{\dagger} & \mathbb{N}(-q)
		\end{pmatrix}, \quad
		\mathbb{M}_4=
		\begin{pmatrix}
			-\mathbb{N}(q) & \mathbb{P} \\
			\mathbb{P}^{\dagger} & -\mathbb{N}(-q)
		\end{pmatrix}, \nonumber \\
		\mathbb{N}(q)=\frac{i}{\sqrt{f(z)}}
		\begin{pmatrix}
			k_y & -\frac{(\omega+q A_t)}{\sqrt{f(z)\chi(z)}}+k_x \\
			\frac{(\omega+q A_t)}{\sqrt{f(z)\chi(z)}}+k_x & -k_y
		\end{pmatrix}, \quad
		\mathbb{P}=
		\begin{pmatrix}
			-i \Phi (z) & 0 \\
			0 & -i \Phi (z)
		\end{pmatrix}.
	\end{gather}
	Because $\xi^{(S)}$ and $\xi^{(C)}$ consist of 4 components, there are 4-independent solutions. We can express a general solution as a combination of these with coefficients $c_{i}, i=1,2,3,4$. For example, if we denote 4 solutions for $\xi^{(S)}$ by $\xi^{(S,i)}, i=1,\cdots ,4$, then arbitrary $\xi^{(S)}$ can be expressed as 
	\begin{align}
		\xi^{(S)}=\sum_{i=1}^{4} c_{i} \xi^{(S,i)}=\mathbb{S}(z)c,
	\end{align}
	where $ \mathbb{S}(z)$ is the 4 by 4 matrix whose $i$-th column is given by $\xi^{(S,i)}$, and $c$ is a column vector whose $i$-th component is $c_{i}$. Similar expression is available for $\xi^{(C)}$. Furthermore, from \eqref{eq:xseq} and \eqref{eq:xceq}, $\xi^{(C)}$ can be expressed in terms of $\xi^{(S)}$, therefore we should use the same $c_{i}$ for $\xi^{(C)}$ also. Namely,
	\begin{align}
		\xi^{(S)}=\mathbb{S}(z)c, \quad \xi^{(C)}=\mathbb{C}(z)c. \label{eq:mrdef}
	\end{align}
	Substituting these to \eqref{eq:xseq} and \eqref{eq:xceq}, we find:
	\begin{align}
		\partial_z\mathbb{S}(z)+\mathbb{M}_1\mathbb{S}(z)+\mathbb{M}_2\mathbb{C}(z)=0, \label{eq:mEQ1} \\
		\partial_z\mathbb{C}(z)+\mathbb{M}_3\mathbb{C}(z)+\mathbb{M}_4\mathbb{S}(z)=0, \label{eq:mEQ2}
	\end{align}
	because $c$ is an arbitrary vector in the solution space. 
	Now, we consider the near boundary behavior of $\xi^{(S)}$ and $\xi^{(C)}$ from those of $\zeta=(\zeta_+,\zeta_-)$ which are given by
	\begin{align}
		\zeta_+=Az^m+Bz^{1-m}, \quad \zeta_-=Dz^{-m}+Cz^{1+m}, \label{eq:asysol}
	\end{align}
	where $A$, $B$, $C$, and $D$ are two-component spinors. If $|m|<1/2$, $ A, D$ terms are leading ones. Therefore 
	\begin{gather}
		\zeta\simeq(Az^{m};Dz^{-m}) , \quad \zeta_c \simeq(\tA^* z^{m};\tD^* z^{-m}). \label{eq:zsb}
 	\end{gather}
	From eq. \eqref{eq:xidef}, 
	\begin{align}
		\xi^{(S)}\simeq(Az^{m}; \tD^* z^{-m}), \quad \xi^{(C)}\simeq(Dz^{-m}; \tA^* z^{m}).
	\end{align}
	Therefore, if we define 
	\begin{align}
		U(z)={\rm diag}(z^m,z^m,z^{-m},z^{-m}),
	\end{align}
	then the near boundary behavior of 
	$\xi^{(S)}$ and $\xi^{(C)}$ can be written as
	\begin{align}
		\xi^{(S)}&=\mathbb{S}(z)c \simeq U(z)\mathbb{S}_0c, \label{eq:ns} \\
		\xi^{(C)}&=\mathbb{C}(z) c\simeq U(z)^{-1}\mathbb{C}_0c, \label{eq:nc} 
	\end{align}
	where $ \mathbb{S}_{0}$ is a matrix whose i-th column is given by the coefficients of the leading terms in $\xi^{(S,i)}$, namely, $(A,{\tilde D}^{*})$. Similar description works for $\mathbb{C}_{0}$. Defining 
	\begin{align}
		\mathcal{J}=\mathbb{S}_0c, \quad \mathcal{C}=\mathbb{C}_0c, \label{SCc}
	\end{align}
	we get
	\begin{align}
		\xi^{(S)}\simeq U(z)\mathcal{J}, \quad \xi^{(C)}\simeq U(z)^{-1} \mathcal{C}. \label{eq:nsc}
	\end{align}	
	It is easy to see that 
	\begin{align}
		\mathcal{J}=(A,\tD^*),\quad \mathcal{C}=(D,\tA^*).
	\end{align}
	Since the contribution of the bulk action to the effective action is zero due to the equation of motion, as we mentioned above, the Green function comes from the boundary action \eqref{eq:actionf} only. Then, the boundary action can be written in terms of $\xi^{(S)}$ and $\xi^{(C)}$:
	\begin{align}
		S_{eff}&= i\int_{z=\epsilon} d^3x(\bar{\zeta}\zeta+\bar{\zeta_c}\zeta_c) \nonumber \\
		&= \int_{z=\epsilon} d^3x \xi^{(S)\dagger}\tilde{\Gamma}\xi^{(C)}+h.c \,\,= \int d^3x \mathcal{J}^{\dagger}\tilde{\Gamma}\mathcal{C}+h.c ,\label{eq:befGR}
	\end{align}
	where ${\tilde \Gamma}=\sigma_0\otimes\gamma^t$. From eq. \eqref{SCc}
	\begin{align}
		\mathcal{C}=\mathbb{C}_0\mathbb{S}_0^{-1}\mathcal{J},
	\end{align}
	so that \eqref{eq:befGR} becomes 
	\begin{align}
		S_{eff}=\int d^3x \mathcal{J}^{\dagger}\mathbb{G}_0\mathcal{J}+ h.c,
	\end{align}
	where 
	\begin{align}
		\mathbb{G}_0=\tilde{\Gamma} \mathbb{C}_0 \mathbb{S}_0^{-1}.
	\end{align}
	The above expression of the action identifies $\mathbb{G}_0$ as the desired retarded Green function $G_R$. 
	We now find the equation satisfied by $\mathbb{G}(z)$. If we substitute equations (\ref{eq:mEQ1}-\ref{eq:mEQ2}) into the definition of Green function, $\mathbb{G}_0=\tilde{\Gamma}\mathbb{C}_0\mathbb{S}_0^{-1}$, and take a derivative with respect to $z$,
	\begin{align}
		\partial_z\mathbb{G}(z)&=\partial_z(\tilde{\Gamma}\mathbb{C}(z)\mathbb{S}(z)^{-1}) \nonumber \\
		&=\tilde{\Gamma}(\partial_z\mathbb{C}(z)\mathbb{S}(z)^{-1}-\mathbb{C}\mathbb{S}(z)^{-1}\partial_z\mathbb{S}(z)\mathbb{S}(z)^{-1}) \nonumber \\
		&=-(\tilde{\Gamma}\mathbb{M}_3\tilde{\Gamma}\mathbb{G}(z)+\tilde{\Gamma}\mathbb{M}_4-\mathbb{G}(z)\mathbb{M}_1-\mathbb{G}(z)\mathbb{M}_2\tilde{\Gamma}\mathbb{G}(z)).
	\end{align}
	In the third line, we used (\ref{eq:mEQ1}-\ref{eq:mEQ2}). Then we have 
	\begin{align}
		\partial_z\mathbb{G}(z)+ [\mathbb{M}_1, \mathbb{G}(z)]-\mathbb{G}(z)\mathbb{M}_2\tilde{\Gamma}\mathbb{G}(z)+\tilde{\Gamma}\mathbb{M}_4 =0,
	\end{align}
	the desired flow equation or Riccati equation for our case. Now we express the boundary Green function $\mathbb{G}_0$ in terms of the bulk quantity $\mathbb{G}(z)$ near the AdS boundary. If we substitute the expressions (\ref{eq:ns}-\ref{eq:nc}) into the definition of Green function,
	\begin{align}
		\mathbb{G}(z)&=\tilde{\Gamma}\mathbb{C}(z)\mathbb{S}(z)^{-1} \nonumber \\
		&\simeq \tilde{\Gamma}U(z)^{-1}\mathbb{C}_0\mathbb{S}_0^{-1}U(z)^{-1} \nonumber \\
		&=U(z)^{-1}\mathbb{G}_0U(z)^{-1}.
	\end{align}
	In the third line, we use $\tilde{\Gamma}^2=\mathbb{1}_{4\times4}$ and $\tilde{\Gamma}U(z)^{-1}\tilde{\Gamma}=U(z)^{-1}$. Finally, the boundary Green function $\mathbb{G}_0$ is given by
	\begin{align}
		\mathbb{G}_0=\lim_{z\rightarrow0}U(z)\mathbb{G}(z)U(z). \label{eq:dobg}
	\end{align}	
	
	\subsection{Horizon Value of the Green Function}
	Here we motivate the use of the flow equation by showing the regularity of the $\mathbb{G}(z)$ at the horizon. Let's take an ansatz for the near horizon behavior, $\zeta_{i}=(1-z/z_H)^a\zeta_{i0}$, where $ \zeta_{i0}$ are constants. If we series expand the Dirac equation near the horizon, we can express the leading terms as follows:
	\begin{gather}
		(1-z/z_H)^{a-\frac{3}{4}}[\mathbf{u}_0+(1-z/z_H)^{\frac{1}{2}}\mathbf{u}_{\frac{1}{2}}+(1-z/z_H)\mathbf{u}_1]= {0}, \label{eq:horizon1} \\
		\mathbf{u}_0=\frac{1}{3^{3/4}}
		\begin{pmatrix}
			-3a{\zeta_{10}}-i{\zeta_{40}}\omega z_H \\
			-3a{\zeta_{30}}+i{\zeta_{20}}\omega z_H \\
			3a{\zeta_{20}}-i{\zeta_{30}}\omega z_H \\
			3a{\zeta_{40}}+i{\zeta_{10}}\omega z_H
		\end{pmatrix},
	\end{gather}
	where $\mathbf{u}_{\frac{1}{2}}$ and $\mathbf{u}_1$ are constant vectors. To make a left-hand side zero, we should make the $\mathbf{u}_0=0$.Then, we can get $a$'s values and the relation between ${\zeta_{i0}}$'s. There are two sets of solutions:
	\begin{align}
		a=\pm\frac{i\omega z_H}{3}, \quad {\zeta_{30}} = \pm {\zeta_{20}}, \quad and \quad {\zeta_{40}}=\mp {\zeta_{10}},
	\end{align}
	which can be recast as
	\begin{align}
		\zeta=\begin{cases}
			(1-z/z_H)^{-\frac{i \omega z_H}{3}}({\zeta_{10}},{\zeta_{20}},-{\zeta_{20}},{\zeta_{10}})^T , & \hbox{for the infalling},\\
			(1-z/z_H)^{\frac{i \omega z_H}{3}} ({\zeta_{10}},{\zeta_{20}},{\zeta_{20}},-{\zeta_{10}})^T, & \hbox{for the outgoing}.
		\end{cases}
	\end{align}
	Notice that if $\zeta$ has the infalling condition, $\zeta_{c}$ automatically takes the outgoing condition. From these conditions, we can construct the horizon behavior of $\xi^{(S)}$ and $\xi^{(C)}$:
	\begin{align}
		\xi^{(S)}&=(Z\zeta_{10},Z\zeta_{20},Z^*\tilde{\zeta}_{20}^*,Z^*\tilde{\zeta}_{10}^*)^T, \\
		\xi^{(C)}&=(-Z\zeta_{20},Z\zeta_{10},-Z^*\tilde{\zeta}_{10}^*,Z^*\tilde{\zeta}_{20}^*)^T,	
	\end{align}
	where $Z=(1-z/z_H)^{-\frac{i w z_H}{3}}$. Then by choosing $\zeta_{i0}$ and $\tilde{\zeta}_{i0}^*$ appropriately, 
	\begin{align}
		\mathbb{S}(z)\simeq
		\begin{pmatrix}
			Z & Z & Z & Z \\
			Z & -Z & Z & Z \\
			Z^{*} & Z^{*} & -Z^{*} & Z^{*} \\
			Z^{*} & Z^{*} & Z^{*} & -Z^{*}
		\end{pmatrix}, \qquad
		\mathbb{C}(z)\simeq
		\begin{pmatrix}
			-Z & Z & -Z & -Z \\
			Z & Z & Z & Z \\
			-Z^{*} & -Z^{*} & -Z^{*} & -Z^{*} \\
			Z^{*} & Z^{*} & -Z^{*} & Z^{*} \\
		\end{pmatrix}.
	\end{align}
	Using $Z^{*}=Z^{-1}$, the horizon value of the matrix Green function is given by
	\begin{align}
		\mathbb{G}(z)=\tilde{\Gamma}\mathbb{C}(z)\mathbb{S}(z)^{-1}=i \mathbb{1}_{4\times4},
	\end{align}
	which is rather surprising: $\mathbb{G}(z)$ is constant near the horizon while $\mathbb{S}(z)$ and $\mathbb{C}(z)$ are singular at $z=z_{H}$. This result is significant in a numerical calculation and is the main reason we want to have the flow equations.
	
	\section{Spectral Density }		
	We now present our result of numerical calculations and then provide an exact result to see some of the analytic details.
	
	\subsection{Comparison between Backreaction and Probe limit}
	In this section, we compare the results of backreaction and the proble limit. First, we mention the method of the backreacted system. To solve the background equations, we write equations of motion as follows:
	\begin{align}
		\Phi''+(\frac{2f'\chi+f\chi'}{2f\chi}-\frac{2}{z})\Phi'+(\frac{Q^2A_t^2}{f^2\chi}+\frac{2}{z^2f})\Phi=0, \label{eq:beqi} \\
		A_t''-\frac{\chi'}{2\chi}A_t'-\frac{2Q^2\Phi^2}{z^2f}=0, \\
		\chi'+z\Phi'^2\chi+\frac{Q^2zA_t^2\Phi^2}{f^2}=0, \\
		f'+(\frac{\chi'}{2\chi}-\frac{3}{z})f-\frac{z^3A_t'^2}{4\chi}+\frac{\Phi^2}{z}+\frac{3}{z}=0, \label{eq:beqf}
	\end{align}
	where $m_\Phi^2=-2$ and $Q$ is charge of the scalar field. And we use a shooting method with horizon behaviors up to the fifth order.
	\begin{align}
		\left(\,A_t(z),\Phi(z),f(z),\chi(z)\,\right) \simeq \sum_{i=0}^{5}(A_{ti},\Phi_{i},f_{i},\chi_{i})\left(1-\frac{z}{z_H}\right)^{i}.
	\end{align}
	One can find relations among coefficient $(A_{ti},\Phi_i,f_i,\chi_i)$ in terms of $(A_{t1}, \Phi_0, \chi_0,z_H)$ by putting again to the above equations (\ref{eq:beqi}-\ref{eq:beqf}). On the other hand, asymptotic behavior near the boundary is given by
	\begin{align}
		A_t(z) &\simeq \mu - \rho z, \\
		\Phi(z) &\simeq \Phi_+z+\langle O \rangle z^2, \label{eq:bbc}
	\end{align}
	where $\rho$ is a charge density, $\Phi_+$ is a scalar source and $\langle O \rangle$ is a scalar codensation. Now, we have two 2nd-order ODE and two 1st-order ODE. It means that we need six boundary conditions whose values are written as
	\begin{align}
		&A_t(z_H)=0, \quad f(z_H)=0, \quad T=control \, parameter, \nonumber \\
		&\mu=control \, parameter, \quad \chi(0)=1, \quad \Phi_+=0.
	\end{align} 
	Now, we have all ingredients to solve the bosonic equations. So, we summarize the process of getting the backreacted solution and finish this section. Input a desired $(T,\mu)=(T_0,\mu_0)$ value and define hyper-plane which is parameterized by $(A_{t1},\Phi_0,\chi_0,z_H)$. Then, one can find the solution of $(A_{t1},\Phi_0,\chi_0,z_H)$, which satisfies $(T,\mu)=(T_0,\mu_0)$. From the calculated value of $(A_{t1},\Phi_0,\chi_0,z_H)$, one can get a configuration of $(A_{t}(z),\Phi(z),f(z),\chi(z))$.
	
\begin{figure}[H]
		\centering 
		\captionsetup{justification=centering}
		\subfloat[AdS-Schwarzschild  ]
		{\includegraphics[width=4.5cm]{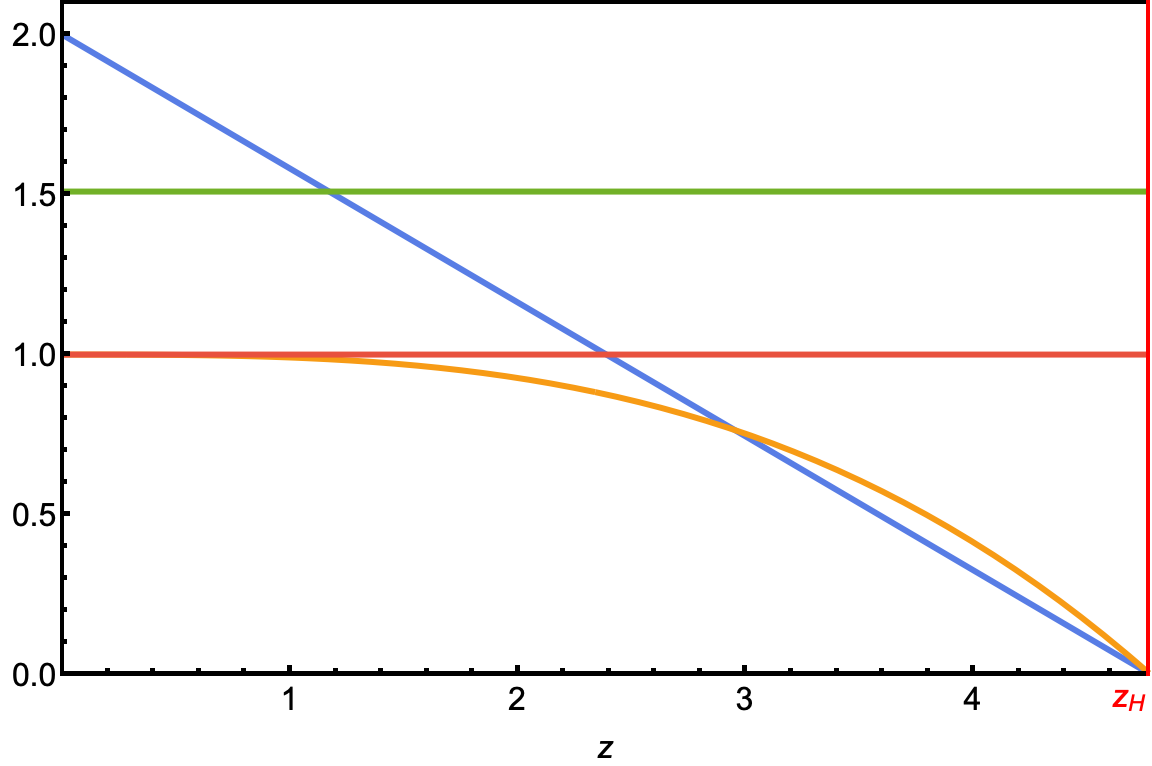}} \hskip 0.01cm	
		\subfloat[RN-AdS  ]
		{\includegraphics[width=4.5cm]{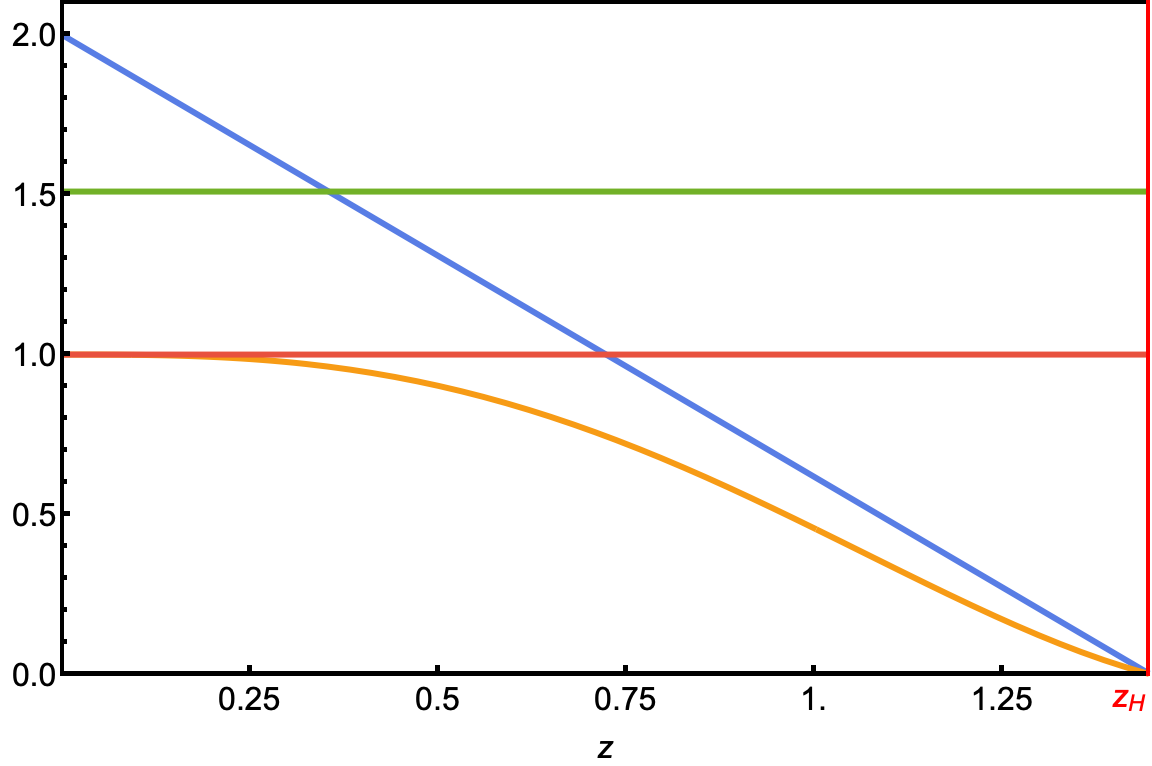}} \hskip 0.01cm
		\subfloat[Backreacted solution]
		{\includegraphics[width=4.5cm]{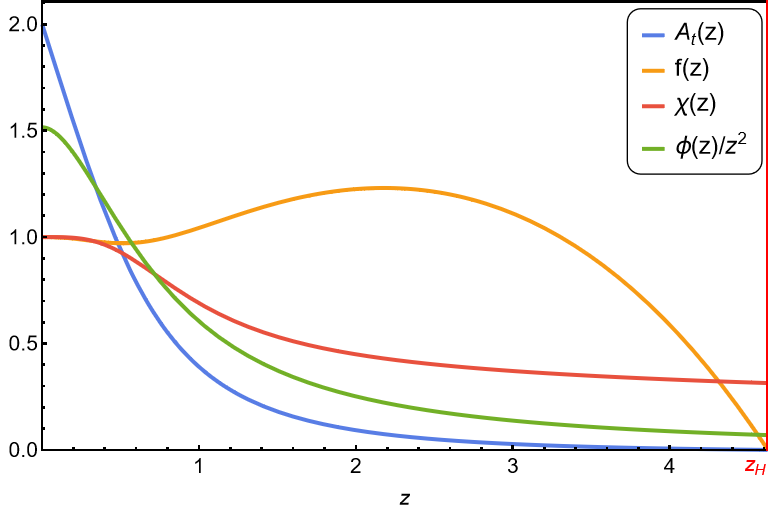}} \\
		\subfloat[AdS-Schwarzschild]
		{\includegraphics[width=4cm]{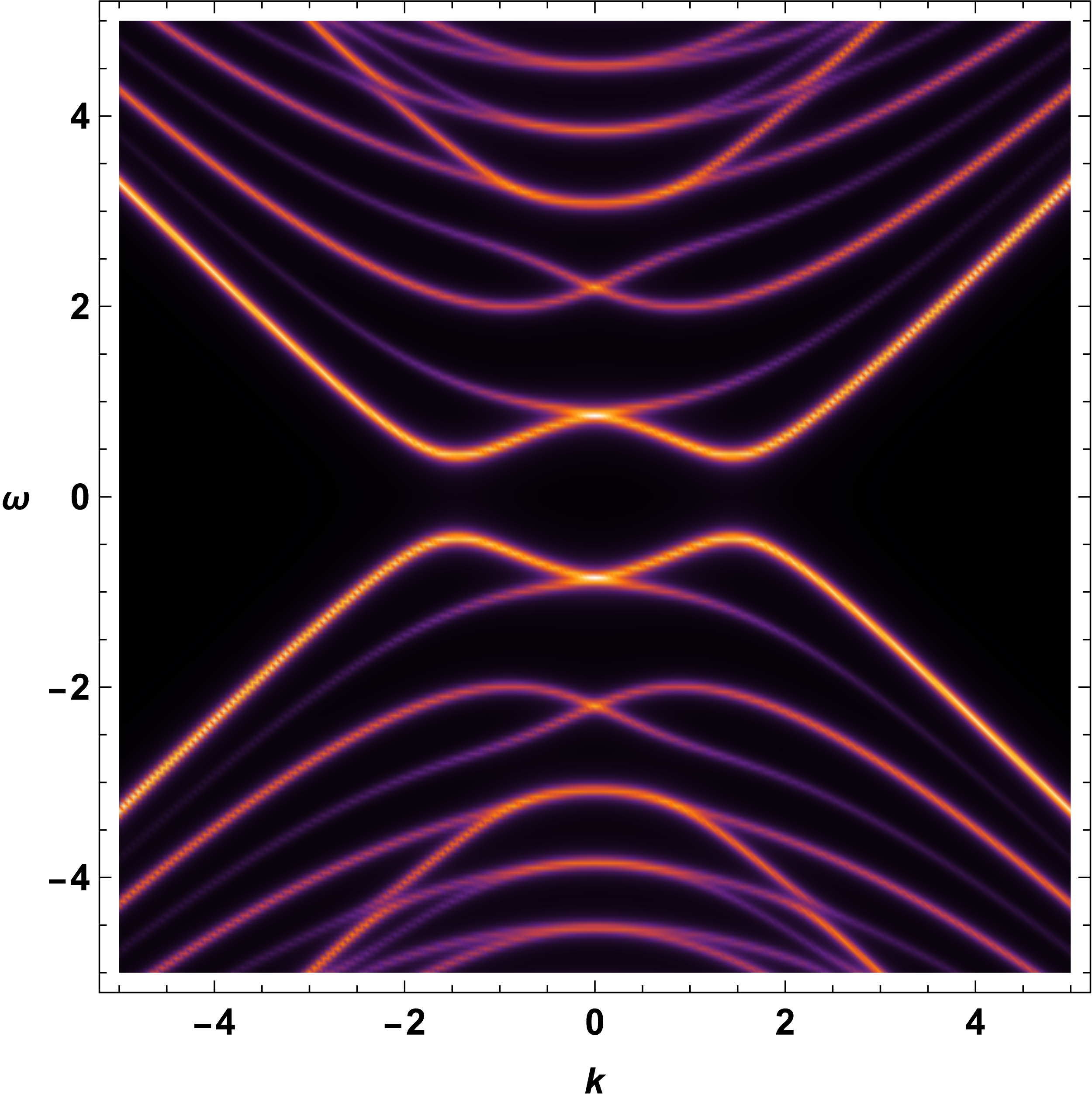}} \hskip 0.8cm
		\subfloat[RN-AdS  ]
		{\includegraphics[width=4cm]{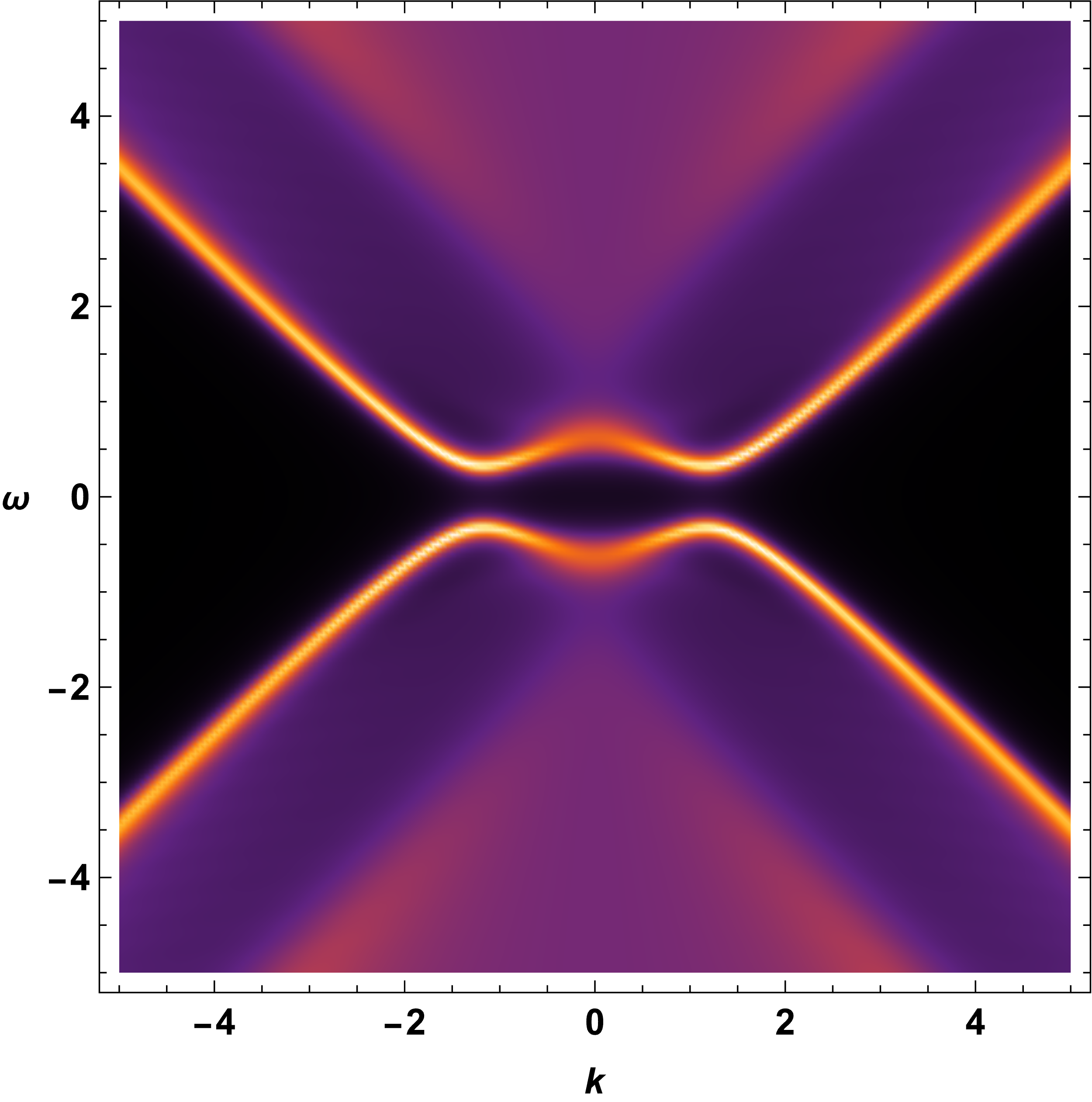}} \hskip 0.8cm
		\subfloat[Backreacted solution]
		{\includegraphics[width=4cm]{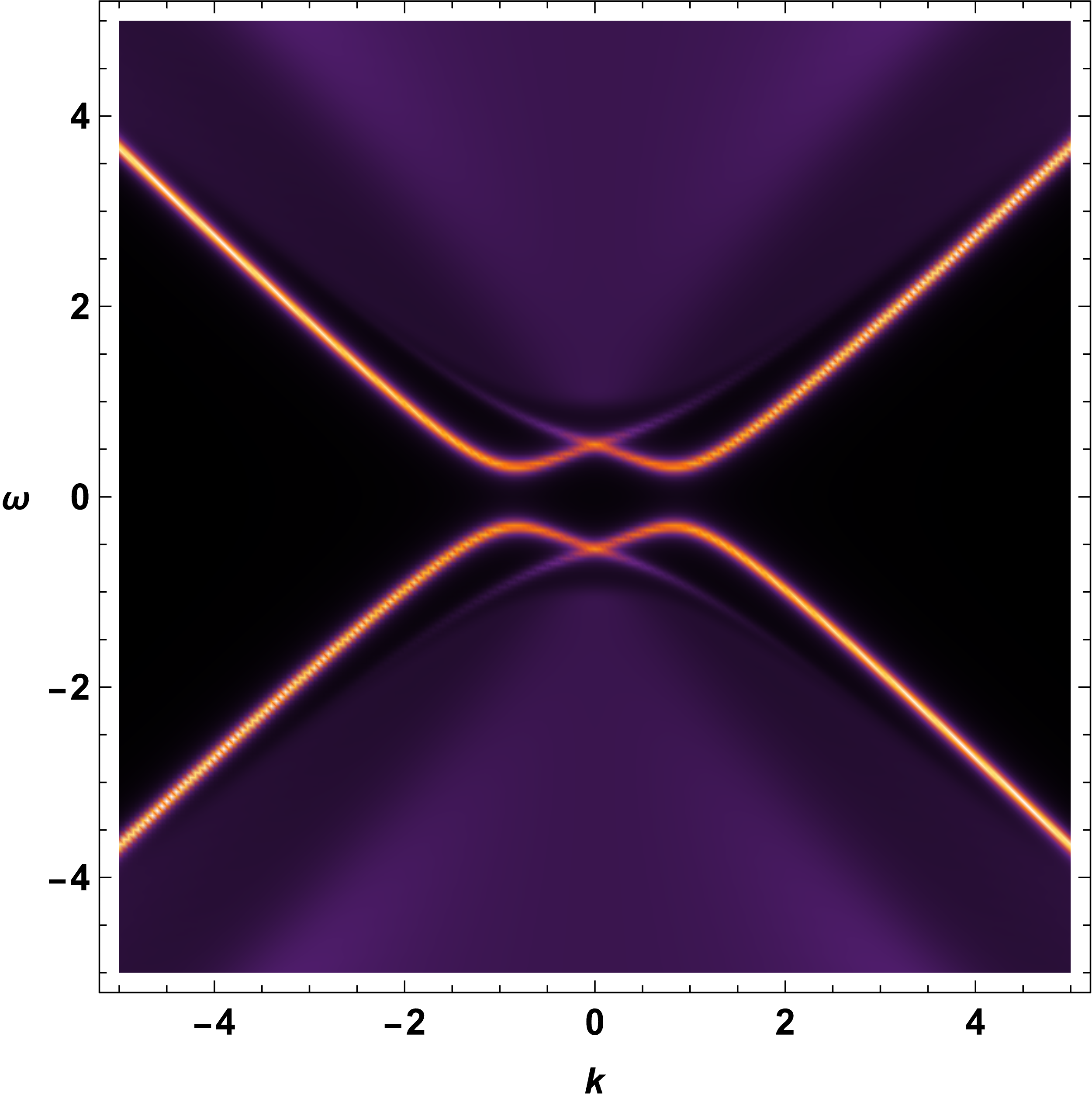}}
		\caption{Background boson fields and Spectral Density(SD). The red vertical line of (a), (b), and (c) means the position of the horizon. We used $T=0.05$, $\mu=1$ and $\langle O \rangle = 1.51$.}
		\label{fig:BPcomp}
	\end{figure}

	For probe limit, we can take two metric solutions, SAdS(Schwarzschild-AdS) and RN-AdS. However, the proper probe limit for holographic superconductors is SAdS\cite{Hartnoll:2020fhc}. Nevertheless, we show the result of RN-AdS for the toy model. We set the same temperature, chemical potential, and scalar condensation for consistency. The boson configuration and spectral density are shown in Figure \ref{fig:BPcomp}.
	Now we discuss the difference between probe limit and back reacted solution. From the boson configuration of figure \ref{fig:BPcomp}, one can notice that the slope of $A_t(z)$(blue curve) near the boundary is higher in (c) than those of (a) and (b). It means the charge density is higher than the probe limits and it is reflected in spectral density(SD) (d,e,f). Comparing SD's, we also see that  \ref{fig:BPcomp}(e) is fuzzier than (d). We suggest that the reason is related to the size of the horizon radius. Small horizon radius means larger black hole size. And the effect of black hole is the scrambling system. Both temperature and density increases the black hole radius as one can see from 
	\begin{gather}
 	r_H=1/ z_{H}=\frac{4 \pi T} {3} \quad \text{for Schwarzschild-AdS}, \\
 		r_H=1/z_{H}=\frac{4\pi T + \sqrt{16 \pi^2 T^2+ 3\mu^2}} {6} \quad \text{for RN-AdS}.
 	\end{gather}
It is natural that the  information scrambling power is higher in larger size black hole so that such black hole should have a fuzzier spectrum. 
 

	\subsection{The dependence of the gap on the chemical potential $\mu$}	
	To understand the effect of the chemical potential, we compare our model with a standard mean-field model of superconductivity \cite{coleman_2015}. The Hamiltonian of the BCS theory model is also that of the action (2.9): 
	\begin{align}
		H=\sum_{k\sigma}\epsilon_kc^{\dagger}_{k\sigma}c_{k\sigma}+\sum_{k}[\Delta^*c_{-k\downarrow}c_{k\uparrow}+\Delta c^{\dagger}_{k\uparrow}c^{\dagger}_{-k\downarrow}],
	\end{align}
	where $\epsilon_k=\frac{\hbar^2k^2}{2m}-\mu$. The energy eigenvalues of this model are given by 
	\begin{align}
		E=\pm\sqrt{(\frac{\hbar^2k^2}{2m}-\mu)^2+|\Delta|^2}, \label{gapmu}
	\end{align}
	which clearly shows particle-hole symmetry as well as the role of the condensation $\Delta$ in gapping the fermion spectrum. Particle and hole bands cross each other, and hybridization gives the avoided crossing, which also generates the gap simultaneously.
	\begin{figure}[H]
	\centering 	
	\captionsetup{justification=centering}
	\subfloat[BCS model]{\includegraphics[width=4cm]{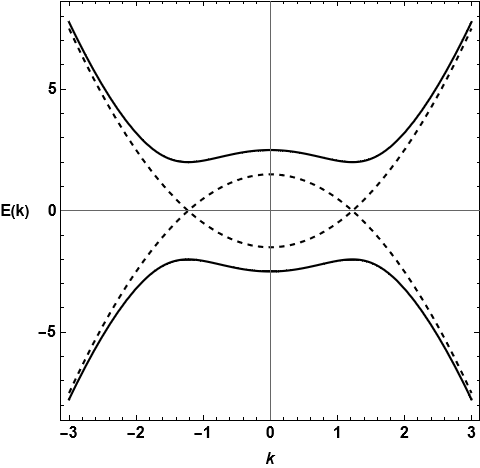}} \hskip 2cm 
	\subfloat[Holographic model]{\includegraphics[width=4cm]{./figure/fermion_back}} 
	\caption{Fermion gap generation in superconductor. (a) BCS model with spectrum given in eq. \eqref{gapmu}. (b) Holographic model with $T=0.05$ and $\mu=1$} \label{fig:meanfd}
	\end{figure}
	See the figure \ref{fig:meanfd}(a). As we can see in the figure \ref{fig:meanfd}(b), there are two effects of chemical potential.
 The first one is to create two minimum(maximum) in the particle(hole) spectrum. This   is a consequence of the downshift of the particle bands and upshift of the hole bands to cross each other, followed by the hybridization of two bands to avoid the crossing. The same effect appears in the BCS mean field theory. 
 The second effect of $\mu$ is to make the spectrum fuzzier, which was  understood in the previous section due to the   increase of black hole radius by the charge density.
 Notice that the fuzziness coming from the density effect is a   rule not a accidental, and this is a characteristic feature of a strongly coupled  and entangled system.

	\subsection{A Solvable model of superconductivity}
\begin{figure}[H]
	\centering 
	\captionsetup{justification=centering}
	\subfloat[SD of ${\rm Tr}G$]
	{\includegraphics[width=4cm]{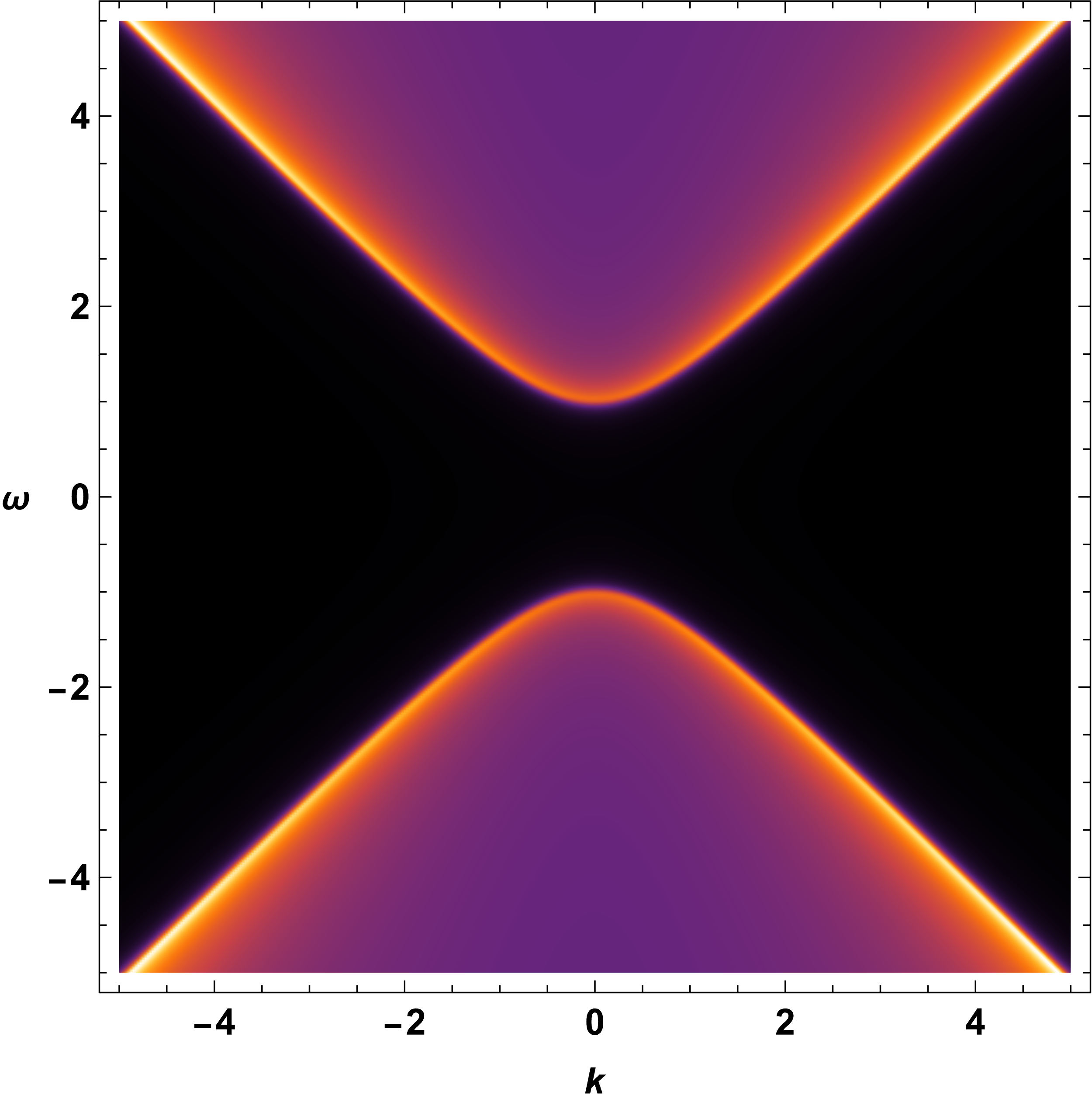}} \hskip 0.5cm
	\subfloat[SD of $TrG$ at $k=0$ line]
	{\includegraphics[width=6.4cm]{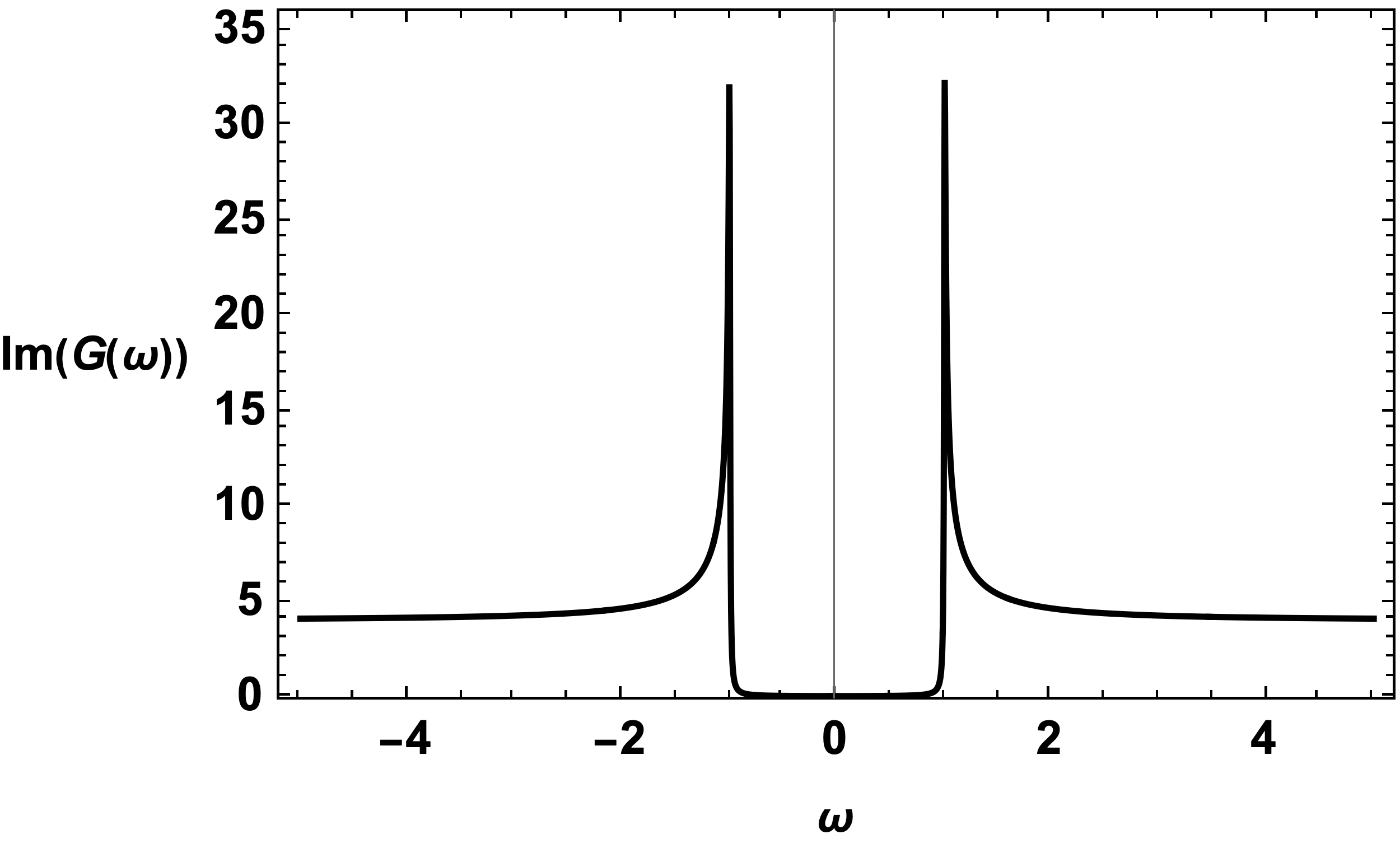}}
	\caption{SD with $\Delta=1$. (a) is a trace of Green's function and (b) is a it's $k=0$ line.}
	\label{fig:SDfig}
\end{figure}

Here, instead of the coupling betweeb the fermion and the scalar field, we consider the case where $\Gamma^z$ comes with a constant $\Delta$ which is analogous to the mass term in flat space. However, the usual bulk mass in AdS does not play the role of the gap creation, while it can do here. The analytic Green's function for such coupling is available at the zero temperature and for the zero bulk mass. The action is given by
\begin{align}
	S_{int}=& \int d^4x \sqrt{-g}\left(\Delta\bar{\psi}\Gamma^{z}\psi_c+h.c. \right).
\end{align}
Notice that  $\Gamma^z$ comes with the vielbein, $\Gamma^z=zf(z)\Gamma^{\underbar{z}}$, so that 
this term  is not the bulk fermion mass. 
Here we present only the result leaving the details in the appendix. The result is given by
\begin{align}
	G_R(\omega,k_x,k_y)=\frac{1}{\sqrt{ k^2+|\Delta|^2-\omega^2}}
	\begin{pmatrix}
		\omega+k_x & -k_y & 0 & \Delta \\
		-k_y & w-k_x & -\Delta & 0 \\
		0 & -\Delta^* & \omega+k_x & -k_y \\
		\Delta^* & 0 & -k_y & \omega-k_x \label{eq:complexgr0}
	\end{pmatrix},
\end{align}
where $G_R=\mathbb{G}_0$ and $k^{2}=k_x^2+k_y^2$.  Notice  that the structure of \eqref{eq:complexgr0} is similar to that of the Quantum Field Theory(QFT) \cite{coleman_2015}. Upper and lower block diagonal matrices represent particle and hole's Green function. And  off-diagonal block matrices represent the pairing condensation parts that result in the gap of the fermion Green function. By diagonalizing it, we can get the quasi-particle Green function.
\begin{align}
	\tilde{G}_R(\omega,k_x,k_y)&={\rm diag}(G_+,G_+,G_-,G_-), \\
	G_+&=\frac{\omega+\sqrt{ k^2+|\Delta|^2}}{\sqrt{ k^2+|\Delta|^2-\omega^2}}, \\
	G_-&=\frac{\omega-\sqrt{ k^2+|\Delta|^2}}{\sqrt{ k^2+|\Delta|^2-\omega^2}}.
\end{align}
Now, we can draw a spectral density (SD) using the definition $\rho={\rm Im}G_R$. The result is shown in Figure \ref{fig:SDfig}.
Now, let's focus on comparing between Holographic and QFT calculations.
\begin{align}
	{\rm Tr} G_R^{Holo}&=\frac{4\omega}{\sqrt{ k^2+\Delta^2-\omega^2}}, \label{eq:gr}\\
	{\rm Tr} G_{R}^{QFT}&=\frac{2\omega}{\omega^2-k_x^2-k_y^2-|\Delta|^2}. \label{eq:grc}
\end{align}
Notice that \eqref{eq:gr} is holographic result, while \eqref{eq:grc} is calculated by QFT\cite{coleman_2015}. 

\begin{figure}[H]	
	\centering		\captionsetup{justification=centering}
	\subfloat[$\mu$=0, $\Delta_0$=1]{\includegraphics[width=4cm]{./figure/SDfig}}
	\subfloat[$\mu$=1, $\Delta_0$=1]{\includegraphics[width=4cm]{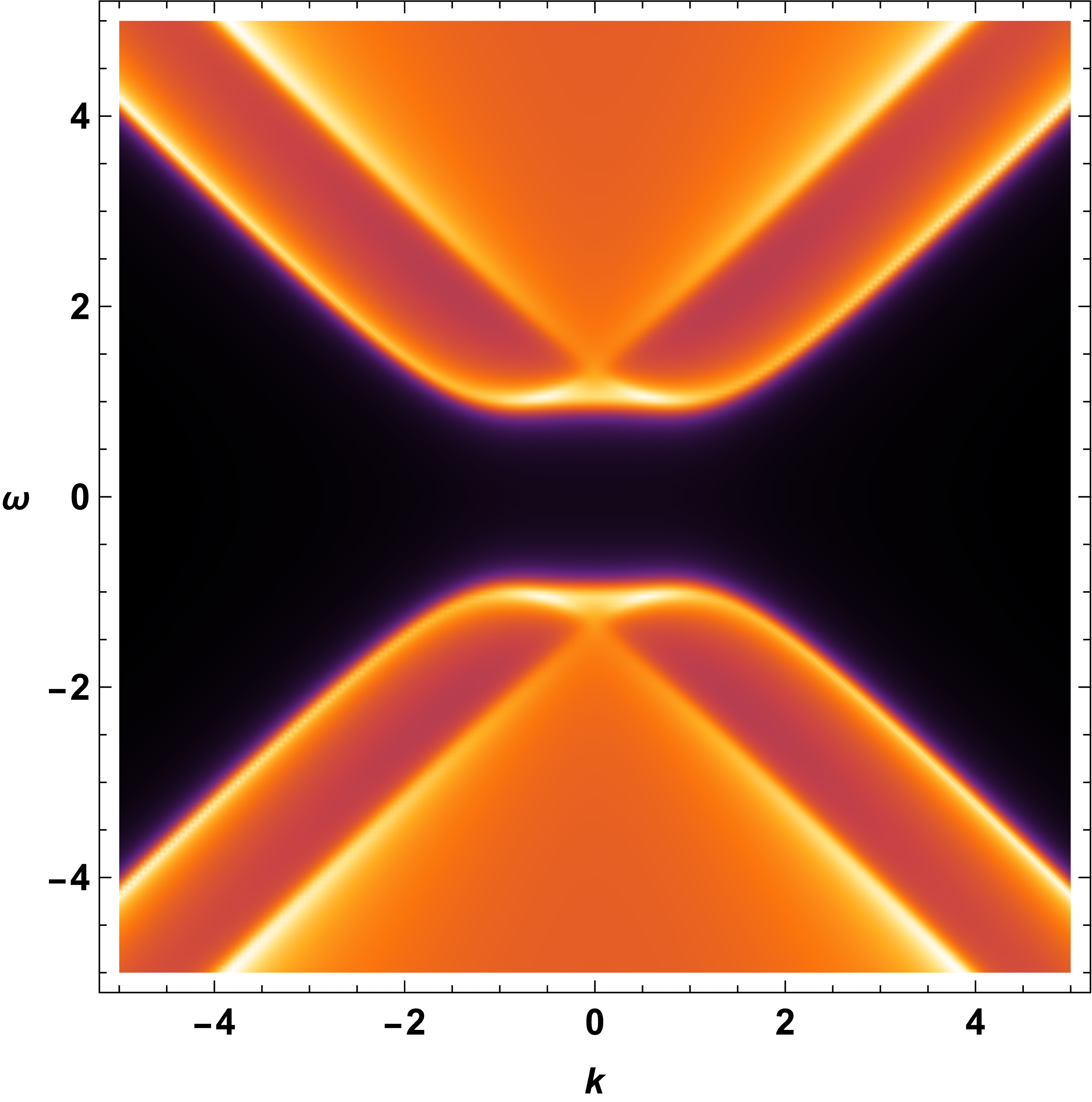}}
	\caption{Spectral density for Fermion coupled with   $\Delta \Gamma^z$. We used $T=0.01$ and $q=1$.} \label{fig:0fig}
\end{figure}

Finally, if we consider the case that both $A_t=\mu(1-z/z_H)$ and $ \Delta $ exist, we can see the effect of the chemical potential, shown in Figure \ref{fig:0fig} by numerical method. 

	\subsection{Comparing the spectra between the four Scalar type interactions}
\begin{figure}[H]
		\centering
		\captionsetup{justification=centering}
		\subfloat[Probe $i\mathbb{1}$]
		{\includegraphics[width=4.cm]{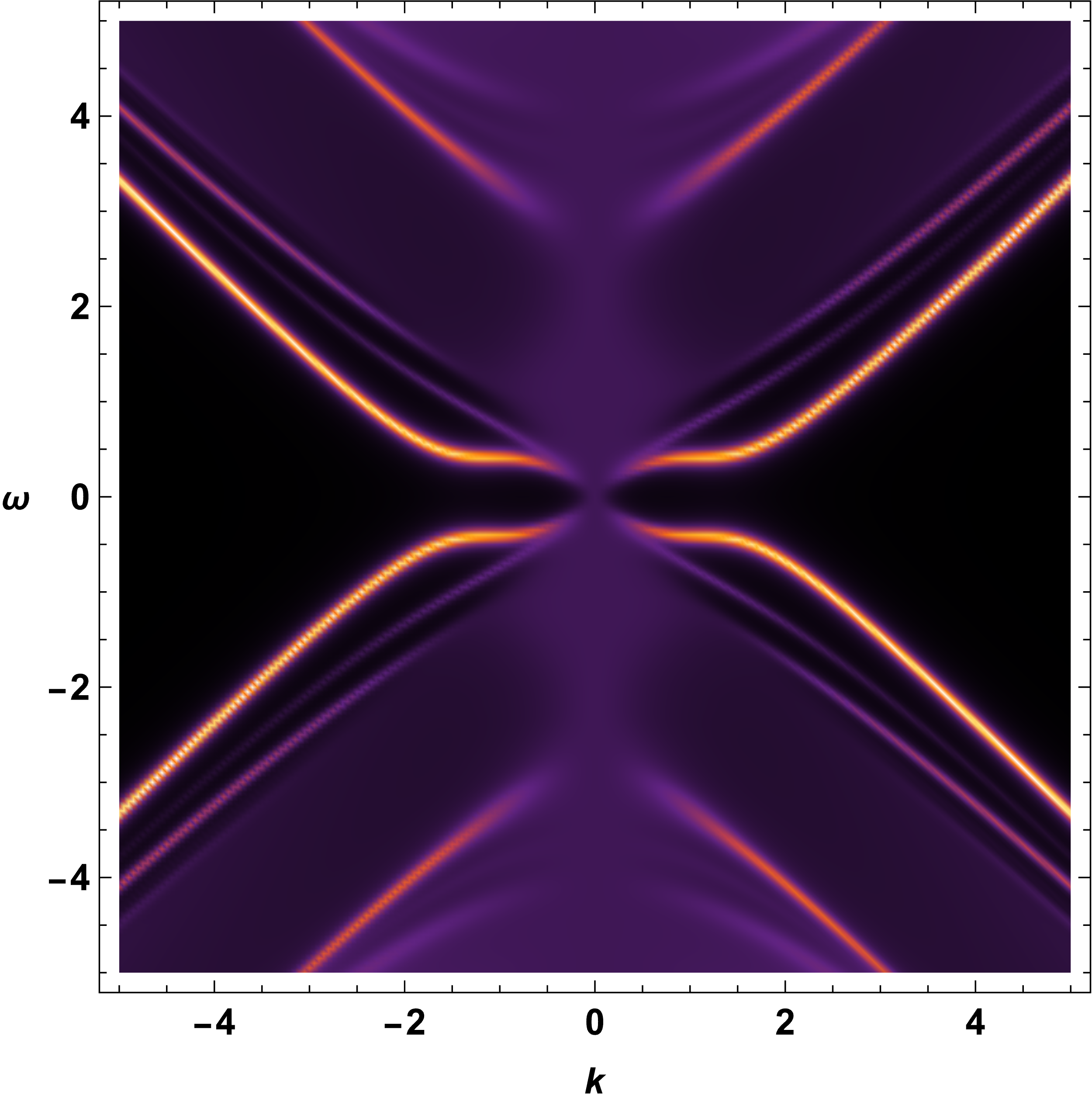}} \hskip 0.5cm 
		\subfloat[Probe $\Gamma^5$]
		{\includegraphics[width=4.cm]{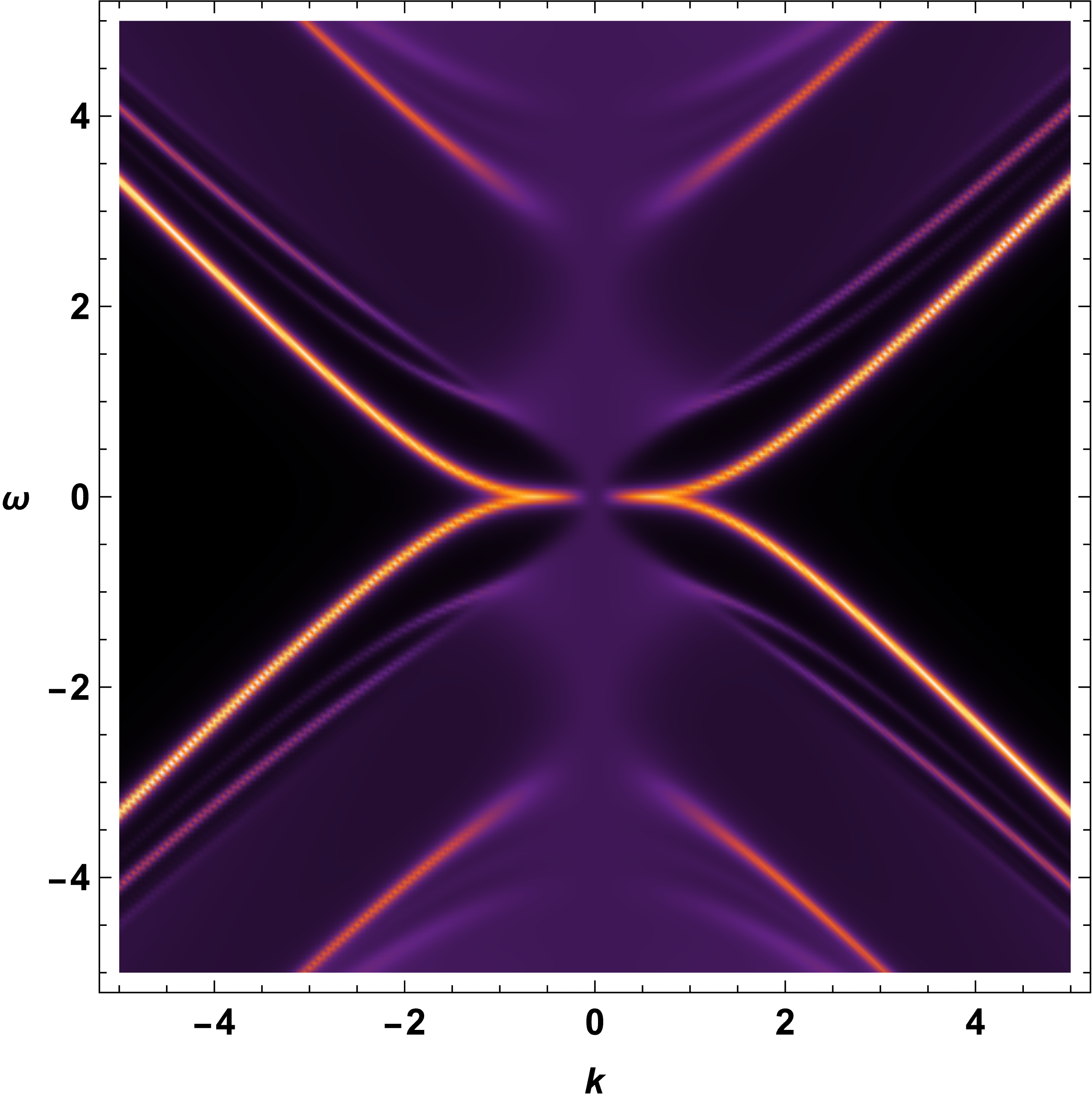}} \\
		\subfloat[Backreacted $i\mathbb{1}$]
		{\includegraphics[width=4.cm]{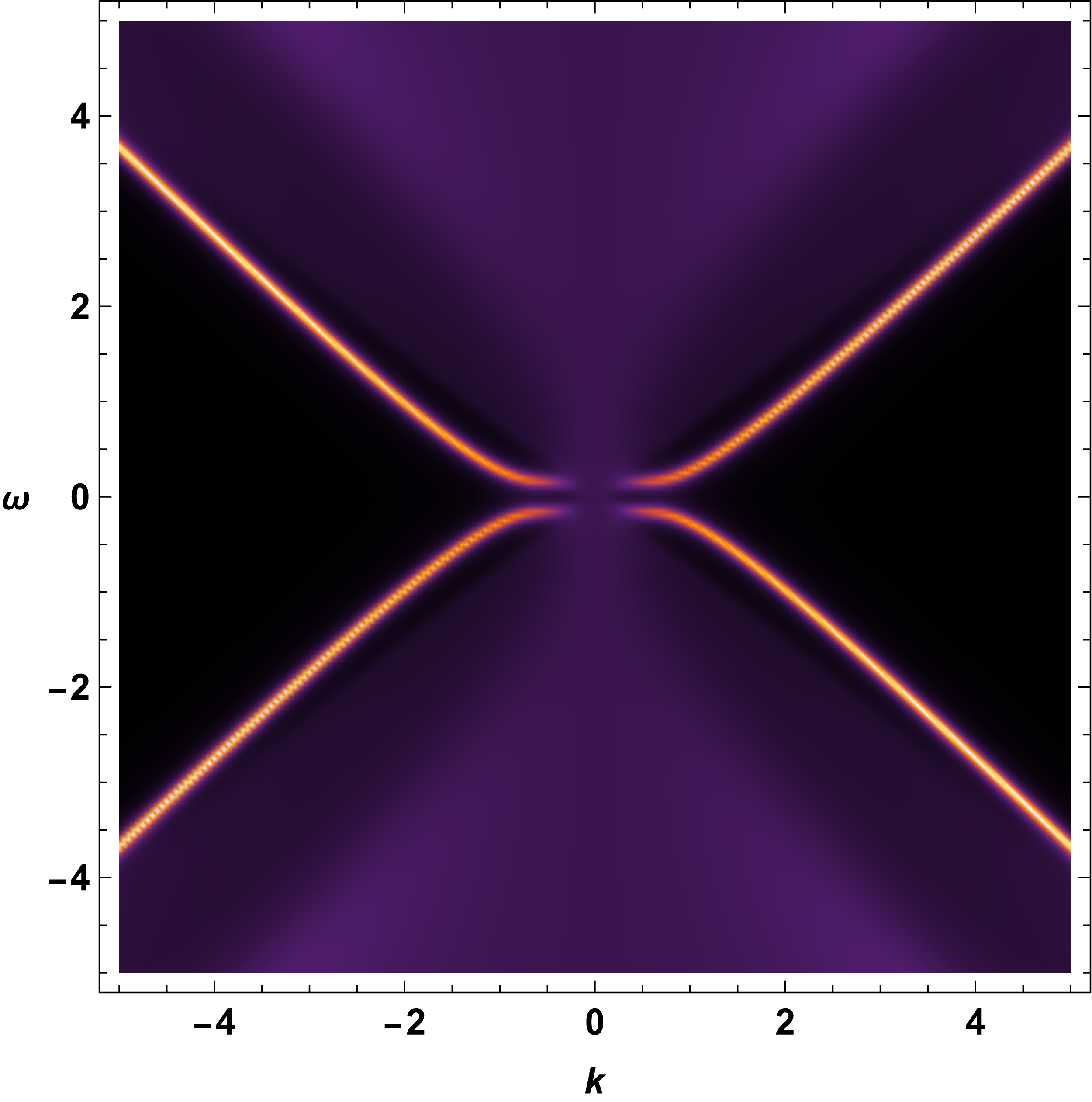}} \hskip 0.5cm 
		\subfloat[Backreacted $\Gamma^5$]
		{\includegraphics[width=4.cm]{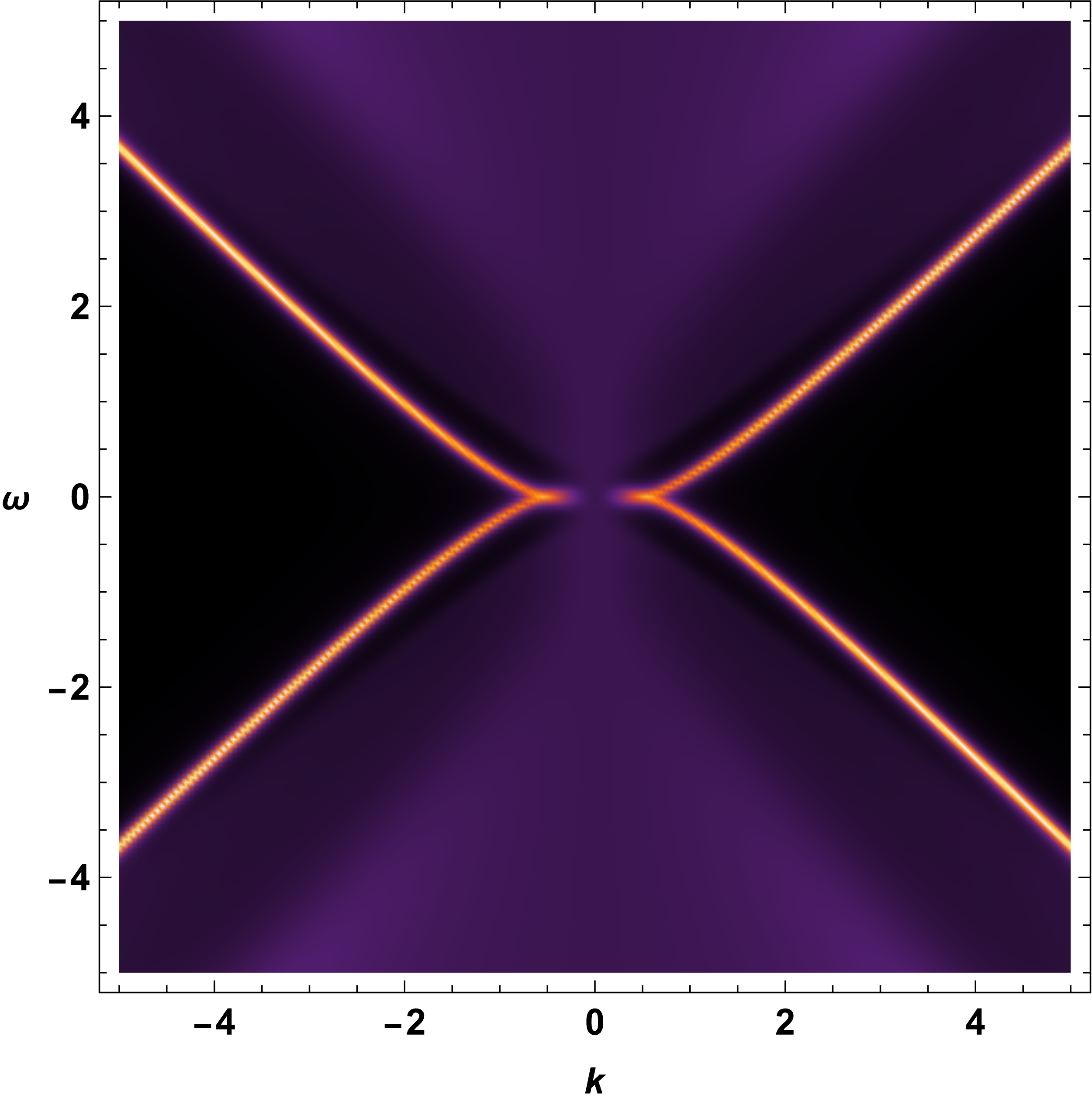}}
		\caption{SD for $\Gamma=i\mathbb{1}$ (a) and (c), and for $\Gamma^{5}$ (b) and (d). Any figure from (a) to (d) does not give the Superconducting gap. Like the $\Gamma^z$, probe limit and backreacted results do not show much difference. We used $T=0.05$, $\mu=1$ and $\langle O \rangle = 1.51$.} \label{fig:others}
	\end{figure}

	In the paper\cite{Oh:2020cym}, we classified the interaction types according to the $\Gamma$ matrix in the interaction term $\Phi {\bar\psi} {\Gamma}\psi$. There can be four classes of scalar type: $i\mathbb{1}, \Gamma^{5}, \Gamma^{z},$ and $\Gamma^{5z}$. It turns out that $i\mathbb{1}, \Gamma^{5},$ give gaps, while the others do not. For superconducting theories   the interaction term should be of the Majorana coupling type, yet, we   have the four scalar types too:
	\begin{align}
		\Phi ^*\bar{\psi_c}\Gamma \psi +h.c, \quad \Gamma=i\mathbb{1}, \Gamma^{5}, \Gamma^{z}, \Gamma^{5z}. \label{majoranaint}
	\end{align}
However, to our surprise,   neither $\mathbb{1}$ nor $\Gamma^5$ type interaction showed the gap in this case. See Figure \ref{fig:others}. Instead, $\Gamma^z$ shows the superconducting gap in the fermion spectrum for Majorana mass type interactions. Therefore, this should be our choice of interaction for the theory of superconductivity. This is the main achievement of this paper from the physical aspect. 
Notice also that spectra in figure \ref{fig:others} are somewhat different from the results of \cite{Faulkner:2009am}.
 
	\section{Discussion}
	In this paper, we reconsidered the fermion spectral function in the presence of the Cooper pair condensation and established the results consistent with the expectations in the s-wave superconducting gap for the first time. We also found that the result of our model is similar to that of the BCS theory in small chemical potential, but very different in higher density cases. We suggested that this is due to the strong correlation between the electrons: one should be reminded that, for a weakly interacting and weakly entangled system, increasing density increases the kinetic energy by the uncertainty principle so that the system becomes a more weakly interacting system because the effective coupling is the ratio of the potential and kinetic energy, i.e., $g_{eff}=V/K$. However, for a highly entangled system where the macroscopic number of particles are entangled, we suspect that the Kinetic energy seems to be frozen, as in the case of the M\"ossbauer effect. Consequently, the effective coupling increases so that one particle spectrum loses its particle character and gives the fuzzy property. 

	
	Finally, we mention a few future directions. Based on figure \ref{fig:BPcomp} and \ref{fig:others}, we will extend the interaction type to the 16 fermion bilinears like the paper \cite{oh2019holographic}. Extension to the p and d-wave superconductor will be interesting to draw a phase diagram whose shape is similar to the cuprate case. It would also be interesting to combine it with a flat band system and investigate its geometric properties.

	\appendix
	 
	\section{Derivation of Analytic Green Function}
	Under the approximations described in section (4.3), the flow equation is given by
	\begin{align}
		\partial_z\mathbb{G}(z)-\mathbb{G}(z)\mathbb{M}_2\tilde{\Gamma}\mathbb{G}(z)+\tilde{\Gamma}\mathbb{M}_4 =0,
	\end{align}
	\begin{align}
\hbox{ with }\quad \quad 		\mathbb{M}_2=
		\begin{pmatrix}
			\mathbb{N}(0) & \mathbb{P} \\
			\mathbb{P}^{\dagger} & \mathbb{N}(0)
		\end{pmatrix},
		\mathbb{M}_4=
		\begin{pmatrix}
			-\mathbb{N}(0) & \mathbb{P} \\
			\mathbb{P}^{\dagger} & -\mathbb{N}(0)
		\end{pmatrix}, \nonumber \\
		\mathbb{N}(q)=i
		\begin{pmatrix}
			k_y & -\omega+k_x \\
			\omega+k_x & -k_y
		\end{pmatrix},
		\mathbb{P}=-i
		\begin{pmatrix}
			\Delta & 0 \\
			0 & \Delta
		\end{pmatrix}.
	\end{align}
	Now, let's take the following ansatz:
	\begin{align}
		\mathbb{G}(z)=
		\begin{pmatrix}
			\mathbb{A}(z) & \mathbb{B}(z) \\
			\mathbb{B}^{\dagger}(z) & \mathbb{A}(z)
		\end{pmatrix}, \quad
		\mathbb{A}=\mathcal{G}(z)
		\begin{pmatrix}
			a_{11} & a_{12} \\
			a_{21} & a_{22}
		\end{pmatrix}, \quad
		\mathbb{B}=\mathcal{F}(z)
		\begin{pmatrix}
			0 & a_{14} \\
			a_{23} & 0
		\end{pmatrix},
	\end{align}
	where each $a_{ij}$ is a complex constant. Then, we can get ten independent differential equations expressed by $\mathcal{G}(z)$ and $\mathcal{F}(z)$:
	\begin{align}
		\cG'(z)+\frac{a_{11}(a_{12}+a_{21})k_y+a_{11}^2(-k_x+\omega)+a_{12}a_{21}(k_x+\omega)}{a_{11}}\cG(z)^2 \nonumber \\
		-Re(a_{14}\Delta^*)\cG(z)\cF(z)+\frac{|a_{14}|^2(k_x+\omega)}{a_{11}}\cF(z)^2+\frac{k_x+\omega}{a_{11}}=0, \label{eq:bareeqi} \\
		\cG'(z)+\frac{(a_{12}^2+a_{11}a_{22})k_y+a_{11}a_{12}(-k_x+\omega)+a_{12}a_{22}(k_x+\omega)}{a_{12}}\cG(z)^2 \nonumber \\
		+(a_{23}^*\Delta-a_{14}\Delta^*)\cG(z)\cF(z)+\frac{a_{14}a_{23}^*k_y}{a_{12}}\cF(z)^2-\frac{k_y}{a_{12}}=0, \\
		\cG'(z)+\frac{(a_{21}^2+a_{11}a_{22})k_y+a_{11}a_{21}(-k_x+\omega)+a_{21}a_{22}(k_x+\omega)}{a_{21}}\cG(z)^2 \nonumber \\
		+(a_{23}\Delta^*-a_{14}^*\Delta)\cG(z)\cF(z)+\frac{a_{23}a_{14}^*k_y}{a_{21}}\cF(z)^2-\frac{k_y}{a_{21}}=0, \\
		\cG'(z)+\frac{a_{22}(a_{12}+a_{21})k_y+a_{22}^2(k_x+\omega)+a_{12}a_{21}(-k_x+\omega)}{a_{22}}\cG(z)^2 \nonumber \\
		+Re(a_{23}\Delta^*)\cG(z)\cF(z)+\frac{|a_{23}|^2(-k_x+\omega)}{a_{22}}\cF(z)^2+\frac{-k_x+\omega}{a_{22}}=0, \\
		\cG'(z)+\frac{a_{11}(a_{12}+a_{21})k_y+a_{11}^2(-k_x+\omega)+a_{12}a_{21}(k_x+\omega)}{a_{11}}\cG(z)^2 \nonumber \\
		+Re(a_{23}\Delta^*)\cG(z)\cF(z)+\frac{|a_{23}|^2(k_x+\omega)}{a_{11}}\cF(z)^2+\frac{k_x+\omega}{a_{11}}=0, \\
		\cG'(z)+\frac{a_{22}(a_{12}+a_{21})k_y+a_{22}^2(k_x+\omega)+a_{12}a_{21}(-k_x+\omega)}{a_{22}}\cG(z)^2 \nonumber \\
		-Re(a_{14}\Delta^*)\cG(z)\cF(z)+\frac{|a_{14}|^2(-k_x+\omega)}{a_{22}}\cF(z)^2+\frac{-k_x+\omega}{a_{22}}=0,
	\end{align}
	\begin{align}		
		\cF'(z)+\frac{(a_{12}^2-a_{11}a_{22})\Delta}{a14}\cG(z)^2+(a_{11}(-k_x+\omega)+a_{22}(k_x+\omega)+2a_{12}k_y)\cG(z)\cF(z) \nonumber \\
		-a_{14}\Delta^*\cF(z)^2+\frac{\Delta}{a_{14}}=0, \\
		\cF'(z)-\frac{(a_{21}^2-a_{11}a_{22})\Delta}{a23}\cG(z)^2+(a_{11}(-k_x+\omega)+a_{22}(k_x+\omega)+2a_{21}k_y)\cG(z)\cF(z) \nonumber \\
		+a_{23}\Delta^*\cF(z)^2-\frac{\Delta}{a_{23}}=0, \\
		\cF'(z)+\frac{(a_{21}^2-a_{11}a_{22})\Delta}{a14^*}\cG(z)^2+(a_{11}(-k_x+\omega)+a_{22}(k_x+\omega)+2a_{12}k_y)\cG(z)\cF(z) \nonumber \\
		-a_{14}^*\Delta\cF(z)^2+\frac{\Delta^*}{a_{14}^*}=0, \\
		\cF'(z)-\frac{(a_{12}^2-a_{11}a_{22})\Delta}{a14}\cG(z)^2+(a_{11}(-k_x+\omega)+a_{22}(k_x+\omega)+2a_{21}k_y)\cG(z)\cF(z) \nonumber \\
		+a_{23}^*\Delta\cF(z)^2-\frac{\Delta^*}{a_{23}^*}=0. \label{eq:bareeqf}
	\end{align}
	Here we can make the coefficients of $\cG(z)$'s and $\cF(z)$'s differential equations be the same. Then, we can find a set of coefficients relation in terms of $a_{11}$ and $a_{14}$:
	\begin{align}
		a_{12}=a_{21}=-\frac{k_y}{k_x+\omega}a_{11}, \quad 
		a_{22}=\frac{-k_x+\omega}{k_x+\omega}a_{11},\quad 
		a_{23}=-a_{14}, \quad a_{14}^*=\frac{\Delta^*}{\Delta}a_{14}. \label{eq:coerel}
	\end{align}
	Now, putting the above relations to the equations (\ref{eq:bareeqi}-\ref{eq:bareeqf}) give a pair of equations as follows:
	\begin{align}
		&\cG'(z)-\frac{a_{11}(k^{2}-\omega^2)}{k_x+\omega}\cG(z)^2-2a_{14}\Delta^*\cG(z)\cF(z)+\frac{a_{14}^2\Delta^*(k_x+\omega)}{a_{11}\Delta}\cF(z)^2+\frac{k_x+\omega}{a_{11}}=0,\nonumber \\
		&\cF'(z)+\frac{a_{11}^2( k^2-\omega^2)}{a_{14}(k_x+\omega)^2}\cG(z)^2-\frac{2a_{11}( k^2-\omega^2)}{k_x+\omega}\cG(z)\cF(z) -a_{14}\Delta^*\cF(z)^2+\frac{\Delta}{a_{14}}=0,
	\end{align}
	where $k^{2}=k_x^2+k_y^2$. \\
	To get a solution of $\mathbb{G}(z)$, we change to the equations of $\mathbb{G}_{ij}$ from those of $\cG(z)$ and $\cF(z)$.
	\begin{align}
		\bG_{11}'(z)-\frac{ k^2-\omega^2}{k_x+\omega}\bG_{11}(z)^2-2\Delta^*\bG_{11}(z)\bG_{14}(z) 
		+\frac{\Delta^*}{\Delta}(k_x+\omega)\bG_{14}(z)^2+k_x+\omega=0, \\
		\bG_{14}'(z)+\frac{ k^2-\omega^2}{(k_x+\omega)^2}\bG_{11}(z)^2-\frac{2( k^2-\omega^2)}{k_x+\omega}\bG_{11}(z)\bG_{14}(z)
		-\Delta^*\bG_{14}(z)^2+\Delta=0.
	\end{align}
	Furthermore, to decouple above equations, we introduce transformation 	\begin{align}
		\begin{pmatrix}
			\bG_{11}(z) \\
			\bG_{14}(z)
		\end{pmatrix}
		=\frac{1}{2}
		\begin{pmatrix}
			1 & 1 \\
			-i\frac{\sqrt{\Delta( k^2-\omega^2)}}{(k_x+\omega)\sqrt{\Delta^*}} & i\frac{\sqrt{\Delta( k^2-\omega^2)}}{(k_x+\omega)\sqrt{\Delta^*}}
		\end{pmatrix}
		\begin{pmatrix}
			h_1(z) \\
			h_2(z)
		\end{pmatrix}. \label{eq:tm}
	\end{align}
	Then, in terms of $h_{1}, h_{2}$, the equations can be expressed as
	\begin{align}
		&h_1'(z)+\frac{-k^2+\omega^2+i|\Delta|\sqrt{ k^2-\omega^2}}{k_x+\omega}h_1(z)^2+\frac{(k_x+\omega)(i|\Delta|+\sqrt{ k^2-\omega^2})}{\sqrt{ k^2-\omega^2}}=0, \\
		&h_2'(z)+\frac{-k^2+\omega^2-i|\Delta|\sqrt{ k^2-\omega^2}}{k_x+\omega}h_2(z)^2+\frac{(k_x+\omega)(-i|\Delta|+\sqrt{ k^2-\omega^2})}{\sqrt{ k^2-\omega^2}}=0.
	\end{align}
	Remarkably, the analytic solution of both equations can be found to be
	\begin{align}
		h_1(z)&=-\frac{(k_x+\omega)(i|\Delta|+\varepsilon_0)}{\varepsilon_0\varepsilon_\Delta}\tanh(\varepsilon_\Delta(z+b_1\varepsilon_0(k_x+\omega))), \\
		h_2(z)&=-\frac{(k_x+\omega)(-i|\Delta|+\varepsilon_0)}{\varepsilon_0\varepsilon_\Delta}\tanh(\varepsilon_\Delta(z+b_2\varepsilon_0(k_x+\omega))),
	\end{align}
	where
	\begin{align}
		\varepsilon_0=\sqrt{ k^2-\omega^2}, \quad \varepsilon_\Delta=\sqrt{ k^2+|\Delta|^2-\omega^2},
	\end{align}
	and $b_1$ and $b_2$ are the constants of integration. From $\bG(z_H)=i\mathbb{1}_{4\times4}$ and \eqref{eq:tm}, the horizon values of $h_1(z_H)$ and $h_2(z_H)$ are given by 
	\begin{align}
		h_1(z)=i, \quad h_2(z)=i,
	\end{align}
	 which in turn request that value of $b_1$ and $b_2$ should be 
	\begin{align}
		b_1=-\frac{1}{\varepsilon_0(k_x+\omega)}\left[\Lambda+\frac{i}{\varepsilon_\Delta}\arctan\left(\frac{\varepsilon_0\varepsilon_\Delta}{(k_x+\omega)(\varepsilon_0+i|\Delta|)}\right)\right], \\
		b_2=-\frac{1}{\varepsilon_0(k_x+\omega)}\left[\Lambda+\frac{i}{\varepsilon_\Delta}\arctan\left(\frac{\varepsilon_0\varepsilon_\Delta}{(k_x+\omega)(\varepsilon_0-i|\Delta|)}\right)\right].
	\end{align}
	Notice that at zero temperature, the horizon can be thought of at $z_H=\infty$. We introduced $\Lambda$ as a cut-off value of $z$. Now, we have all ingredients for the boundary Green function. For the massless fermion, the Green function can be written as
	\begin{align}
		G_R=\lim_{z\rightarrow0}\bG(z).
	\end{align}
	From this, the value of $h_1(z)$ and $h_2(z)$ at the AdS boundary are
	\begin{align}
		h_1(z)=\frac{(k_x+\omega)(i|\Delta|+\varepsilon_0)}{\varepsilon_0\varepsilon_\Delta}, \\
		h_2(z)=\frac{(k_x+\omega)(-i|\Delta|+\varepsilon_0)}{\varepsilon_0\varepsilon_\Delta}.
	\end{align}
	To find value of $G_R$, let's recover the value of $\bG_{11}(0)$ and $\bG_{14}(0)$ using \eqref{eq:tm}:
	\begin{align}
		G_{R,11}&:=\bG_{11}(0)=\frac{k_x+\omega}{\sqrt{ k^2+|\Delta|^2-\omega^2}}, \\
		G_{R,14}&:=\bG_{14}(0)=\frac{\Delta}{\sqrt{ k^2+|\Delta|^2-\omega^2}}.
	\end{align}
	Finally, with the relation \eqref{eq:coerel}, we can get the full-component Green function $G_R$ already described in section (4.3):
	\begin{align}
		G_R(\omega,k_x,k_y)=\frac{1}{\sqrt{ k^2+|\Delta|^2-\omega^2}}
		\begin{pmatrix}
			\omega+k_x & -k_y & 0 & \Delta \\
			-k_y & w-k_x & -\Delta & 0 \\
			0 & -\Delta^* & \omega+k_x & -k_y \\
			\Delta^* & 0 & -k_y & \omega-k_x \label{eq:complexgr}
		\end{pmatrix}.
	\end{align}

	\acknowledgments
	This work is supported by Mid-career Researcher Program through the National Research Foundation of Korea grant No. NRF-2021R1A2B5B02002603. 
	We also thank the APCTP for the hospitality during the focus program, “Quantum Matter and Entanglement with Holography”, where part of this work was discussed.

\bibliographystyle{jhep}
\bibliography{HSC_ref.bib}

\end{document}